\documentclass[preprint,showpacs,aps,pre]{revtex4}
\usepackage{amsmath}
\usepackage{graphicx}
\usepackage{dcolumn}
\usepackage{bm}

\begin{document}

\title{\textbf{Fourier's Law for a Granular Fluid}}
\author{James W. Dufty}
\affiliation{Department of Physics, University of Florida, Gainesville, FL 32611}
\date{\today }

\begin{abstract}
Newton' viscosity law for the momentum flux and Fourier's law for the heat
flux define Navier-Stokes hydrodynamics for a simple, one component fluid.
There is ample evidence that a hydrodynamic description applies as well to a
mesoscopic granular fluid with the same form for Newton's viscosity law.
However, theory predicts a qualitative difference for Fourier's law with an
additional contribution from density gradients even at uniform temperature.
The reasons for the absence of such terms for normal fluids are indicated,
and a related microscopic explanation for their existence in granular fluids
is presented.
\end{abstract}

\maketitle \affiliation{Department of Physics, University of
Florida, Gainesville, FL 32611}

\section{Introduction}

Granular fluids are of increasing interest to the Chemical Engineering,
Physics, and Materials Sciences communities, for different and complementary
reasons. On the practical side are industrial issues of agricultural,
pharmaceutical, and chemical significance for packing and transport of
grains. More recently, the planned expeditions to the moon and Mars require
an understanding of the surface regolith, a new form of granular matter. In
spite of the growing phenomenology for granular fluids, the fundamental
descriptions for the simplest states remain subject to question. An
important class of questions involve the form and conditions for a
hydrodynamic description \cite%
{haff83,campbell90,kadanoff99,Goldhirsch03,Dufty03}. Here, one new feature
of the granular Navier-Stokes hydrodynamic representation is addressed.

There are two peculiarities of the Navier-Stokes equations for granular
fluids relative to those for normal fluids. One is a dissipative source of
energy due to the inelasticity of collisions among the grains, easiliy
understood as a consequence of the mesocopic nature of the constituent
particles. More puzzling is the modification of Fourier's law, with an
additional contribution to the heat flux $\mathbf{s}$ due to a density
gradient%
\begin{equation}
\mathbf{s}=-\lambda \boldsymbol{\nabla }T-\mu \boldsymbol{\nabla }n.
\label{1.1}
\end{equation}%
Here $\lambda $ is the thermal conductivity and $\mu $ is the new transport
coefficient characterizing heat flow in an isothermal, inhomogeneous fluid.
Although physical interpretations of this latter effect have been given \cite%
{Candela05}, it is instructive to associate it with the fundamental
differences between normal and granular fluids. At the continuum level fluid
symmetry alone leads to the general form (\ref{1.1}) in both cases.
Therefore, to better understand the presence of the density gradient
contribution for granular fluids, it is useful to ask why it is absent for
normal fluids.

This question is answered here at the level of the Boltzmann kinetic
equation for a gas and using the more general statistical mechanics of
linear response. In fact, a definitive conclusion requires information
beyond continuum mechanics from the microscopic basis for hydrodynamics. The
hydrodynamic fields are averages of the densities associated with the global
invariants of mass, energy, and momentum. The hydrodynamic equations are
then obtained from representations of the corresponding fluxes in the
average microscopic conservation laws for these densities. There are two
central observations here that are responsible for $\mu =0$ in normal
fluids. The first is that the mass flux is equal to the momentum density.
The second is that the reference state (equilibrium or local equilibrium) is
entirely determined by the invariants. The first implies that there is no
dissipation in the continuity equation for the mass density. The second
imposes important constraints on the representation for transport
coefficients in terms of correlation functions for the fluxes. The most
compact version of these is Onsager's theorem \cite{Onsager}, which is
discussed next. Then more explicit realizations of these correlation
functions are obtained from low density kinetic theory and formally exact
statistical mechanics.

It should be emphasized that the existence of the coefficient $\mu $ is not
in question, only its origin and interpretation. The coefficient has been
determined for idealized isolated cooling granular gases \cite{Brey05}, and
measured in both simulations \cite{Brey04} and experiments \cite{Candela04}
for shaken fluids in a gravitational field

It is a pleasure to dedicate this work to Keith Gubbins whose contributions
to the kinetic theory and statistical mechanics of fluids have influenced me
for more than thirty years. He has been an exceptional combination of
mentor, colleague, role model, and friend to me and many others.

\section{A Consequence of Onsager's Theorem}

\label{sec2} As noted above, the Navier-Stokes conditions of fluid symmetry
and small spatial gradients do not constrain the "constitutive equation" for
the heat flux beyond the form (\ref{1.1}). However, in 1931 Onsager made a
seminal observation relating the transport coefficients of a normal fluid to
the underlying statistical mechanics of the fluid at equilibrium \cite%
{Onsager}. The Onsager regression hypothesis states that on long space and
time scales the decay of spontaneous fluctuations in an equilibrium fluid is
governed by the same laws as for nonequilibrium states displaced slightly
from equilibrium. This is effectively what has become formalized as linear
response theory. The symmetry property of these equilibrium fluctuations
resulting from microscopic time reversal invariance of the dynamics leads to
relations among the transport coefficients. These are the familiar Onsager
relations for a fluid mixture. However, it is less well recognized that
these same relations for a one component fluid imply that $\mu =0$ in Eq. (%
\ref{1.1}) \cite{McL89,dGyM69}

To show this explicitly, consider the exact macroscopic balance equations
for the number density $n(\mathbf{r},t)$, energy density $e(\mathbf{r},t)$,
and the momentum density $\mathbf{g}(\mathbf{r},t)$
\begin{equation}
\partial _{t}n(\mathbf{r},t)+m^{-1}\nabla \cdot \mathbf{g}(\mathbf{r},t)=0,
\label{2.1}
\end{equation}%
\begin{equation}
\partial _{t}e(\mathbf{r},t)+\nabla \cdot \mathbf{s}(\mathbf{r},t)=w(\mathbf{%
r},t),  \label{2.2}
\end{equation}%
\begin{equation}
\partial _{t}g_{i}(\mathbf{r},t)+\partial _{j}t_{ij}\left( \mathbf{r}%
,t\right) =0,  \label{2.3}
\end{equation}%
where $\mathbf{g}(\mathbf{r},t),$ $\mathbf{s}(\mathbf{r},t),$ and $%
t_{ij}\left( \mathbf{r},t\right) $ are the associated mass, energy, and
momentum fluxes and $m$ is the mass. Here, it has already been assumed that
the flux of mass is the same as the momentum density. In fact, this requires
proof from the underlying microscopic balance equations. Also, $w(\mathbf{r}%
,t)$ is an energy source term that could be due to an external force doing
work on the fluid, or the internal collisional energy loss of a granular
fluid. The fluxes have a contribution due to convection and a purely
dissipative contribution that occurs in the local rest frame at each point
of the fluid. This separation is easily identified by a local Galilean
transformation with the results \cite{McL89}%
\begin{equation}
\mathbf{g}(\mathbf{r},t)\equiv mn(\mathbf{r},t)\mathbf{u}(\mathbf{r},t),
\label{2.4}
\end{equation}%
\begin{equation}
s_{i}(\mathbf{r},t)=\left( e(\mathbf{r},t)+p(\mathbf{r},t)\right) u_{i}(%
\mathbf{r},t)+t_{ij}^{\ast }\left( \mathbf{r},t\right) u_{j}(\mathbf{r}%
,t)+s_{i}^{\ast }(\mathbf{r},t)  \label{2.5}
\end{equation}%
\begin{equation}
t_{ij}\left( \mathbf{r},t\right) =p(\mathbf{r},t)\delta _{ij}+mn(\mathbf{r}%
,t)u_{i}(\mathbf{r},t)u_{j}(\mathbf{r},t)+t_{ij}^{\ast }\left( \mathbf{r}%
,t\right) .  \label{2.6}
\end{equation}%
Equation (\ref{2.4}) defines the local flow velocity $\mathbf{u}(\mathbf{r}%
,t)$, while (\ref{2.5}) and (\ref{2.6}) define the irreversible energy flux $%
s_{i}^{\ast }(\mathbf{r},t)$ and momentum flux $t_{ij}^{\ast }\left( \mathbf{%
r},t\right) $ in the rest frame. Finally, $p(\mathbf{r},t)$ is the
hydrostatic pressure which must be specified as a function of the local
energy and density.

The set of exact equations (\ref{2.1})-(\ref{2.6}) become a closed set of
hydrodynamic equations once the "constitutive equations" for $s_{i}^{\ast }(%
\mathbf{r},t)$ and $t_{ij}^{\ast }\left( \mathbf{r},t\right) $ are given in
terms of the number, energy, and momentum densities. In practice, it is
useful to introduce a conjugate set of variables through the change of
variables%
\begin{equation}
\chi _{\alpha }(\mathbf{r},t)\equiv -\frac{\partial s\left( \left\{ y_{\beta
}(\mathbf{r},t)\right\} \right) }{\partial y_{\alpha }(\mathbf{r},t)},%
\hspace{0.2in}y_{\alpha }(\mathbf{r},t)\Leftrightarrow \left( n(\mathbf{r}%
,t),e(\mathbf{r},t),\mathbf{g}(\mathbf{r},t)\right) .  \label{2.7}
\end{equation}%
Here $s\left( \left\{ y_{\beta }(\mathbf{r},t)\right\} \right) $ is the
equilibrium entropy density for the fluid as a function of the number,
energy, and momentum densities at each point of the fluid. For example, it
follows that $\chi _{2}=-1/T$ which defines the temperature as a function of
the density and energy. Note that although this entropy function is defined
for an equilibrium fluid, it is used here simply as the mathematical
generator of a change of variables. As such, it applies even to states far
from equilibrium and also to the granular fluid.

For states near uniform equilibrium, the leading contributions to $\mathbf{s}%
^{\ast }(\mathbf{r},t)$ and $t_{ij}^{\ast }\left( \mathbf{r},t\right) $ are
linear in gradients of the $\left\{ y_{\alpha }\right\} $, or equivalently,
gradients of $\left\{ \chi _{\alpha }\right\} $. The resulting equations (%
\ref{2.1})-(\ref{2.3}) are the Navier-Stokes order hydrodynamics
\begin{equation}
\partial _{t}y_{\alpha }(\mathbf{r},t)+\nabla \cdot \mathbf{j}_{\alpha
}\left( \mathbf{r},t\right) =w_{\alpha },  \label{2.9}
\end{equation}%
where the fluxes are given by%
\begin{equation}
\mathbf{j}_{\alpha }(\mathbf{r},t)=\mathbf{j}_{\alpha }^{(0)}(\mathbf{r},t)+%
\mathbf{j}_{\alpha }^{\ast }(\mathbf{r},t),\hspace{0.2in}\mathbf{j}_{\alpha
}(\mathbf{r},t)\Leftrightarrow \left( \mathbf{g}(\mathbf{r},t),\mathbf{s}(%
\mathbf{r},t),t_{ij}(\mathbf{r},t)\right) .  \label{2.10}
\end{equation}%
Here, $\mathbf{j}_{\alpha }^{(0)}$ are the Euler order contributions
identified from (\ref{2.5}) and (\ref{2.6}), and $\mathbf{j}_{\alpha }^{\ast
}$ are the irreversible contributions. At Navier-Stokes order the latter are
given by the linear constitutive equations \cite{McL89,dGyM69}%
\begin{equation}
\mathbf{j}_{\alpha }^{\ast }(\mathbf{r},t)\mathbf{=}-\sum_{\beta }\mathrm{L}%
_{\alpha \beta }\left( \left\{ y_{\beta }(\mathbf{r},t)\right\} \right) :%
\frac{\partial \chi _{\beta }(\mathbf{r},t)}{\partial \mathbf{r}}.
\label{2.11}
\end{equation}%
The energy flux in this approximation is the generalized Fourier's law (\ref%
{1.1}) and the momentum flux is the usual Newton's viscosity law.

Up to this point only the macroscopic conservation laws, gradient expansion
near the reference homogeneous state, and fluid symmetry have been used. For
a normal fluid Onsager's regression hypothesis allows identification of the
transport coefficients $\mathrm{L}_{\alpha \beta }$ in the form%
\begin{equation}
\mathrm{L}_{\alpha \beta }=\int_{0}^{\infty }dt\mathrm{C}_{\alpha \beta }(t),
\label{2.12}
\end{equation}%
where $\mathrm{C}_{\alpha \beta }(t)$ is an equilibrium time correlation
function for two microscopic fluxes corresponding to the macroscopic $%
\mathbf{j}_{\alpha }$. These fluxes have a definite parity under the
transformation $t\rightarrow -t$, and reversal of all particle velocities $%
\left\{ \mathbf{v}_{i}\right\} \mathbf{\rightarrow }\left\{ \mathbf{-v}%
_{i}\right\} $. Since both the dynamics and the equilibrium ensemble are
invariant under this transformation it follows that
\begin{equation}
\mathrm{C}_{\alpha \beta }(t)=\tau _{\alpha }\tau _{\beta }\mathrm{C}_{\beta
\alpha }(t),  \label{2.13}
\end{equation}%
where $\tau _{\alpha }=\pm 1$, depending on the parity of the associated
flux. Therefore $L_{\alpha \beta }$ has the symmetry%
\begin{equation}
\mathrm{L}_{\alpha \beta }=\tau _{\alpha }\tau _{\beta }\mathrm{L}_{\beta
\alpha }.  \label{2.14}
\end{equation}%
This is Onsager's theorem.

The coefficient $\mu $ in (\ref{1.1}) is identified as%
\begin{equation}
\mu =L_{21}\frac{\partial \chi _{1}}{\partial n}\mid _{T}.  \label{2.14a}
\end{equation}%
An important observation now is that the macroscopic conservation law for
the number density, the continuity equation, has no dissipative
contribution, $L_{1\beta }=0$, since the mass flux is equal to the momentum
density. Consequently, Onsager's theorem gives
\begin{equation}
L_{\beta 1}=L_{1\beta }=0,  \label{2.14b}
\end{equation}%
and Fourier's law becomes
\begin{equation}
\mathbf{s}^{\ast }(\mathbf{r},t)\mathbf{=-}\lambda \left( \left\{ y_{\beta }(%
\mathbf{r},t)\right\} \right) \nabla T(\mathbf{r},t),\hspace{0.2in}\lambda
=L_{22}T^{-2}.  \label{2.15}
\end{equation}%
(There is no coupling to the velocity gradients, $L_{\alpha 3}=0$, from
fluid symmetry). Thus, $\mu =0$ for a normal fluid as a consequence of the
fact that the number flux in the continuity equation has no dissipative
contribution, and the symmetry of the correlation functions resulting from
the fact that the reference state is a function of the invariants. These
properties are demonstrated more explicitly in the further microscopic
elaboration below.

The above analysis holds as well for granular fluids, except for Onsager's
theorem. The condition $L_{1\beta }=0$ still applies but the key symmetry, (%
\ref{2.13}), rests on both the equilibrium Gibbs reference state and the
invariance of the dynamics for the system. Neither the homogeneous reference
state nor the dynamics of a granular fluid have these symmetries, so neither
the usual form of Onsager's theorem nor the conclusion that $\mu =0$ can be
extended to granular fluids. This failure of Onsager's theorem occurs as
well for normal fluids in nonequilibrium stationary states. However, it is
possible that some other symmetry could apply to enforce the usual form of
Fourier's law for granular fluids. Thus, it is important to look in more
detail at the mesoscopic (kinetic theory) and microscopic (statistical
mechanics) representations of this constitutive relation. This is the
objective of the next two sections.

\section{Kinetic theory}

\label{sec4}

The simplest fluid is a low density gas for which the appropriate kinetic
theory is given by the Boltzmann equation. The usual derivations of this
equation for a normal gas can be extended to the granular gas as well, to
account for binary collisions that are inelastic \cite%
{Dufty01,vanN01,Poschel01}. It provides an important testing ground for the
derivation of hydrodynamics and for exploration of conceptual issues as
well. The application to a normal gas is considered first, and then its
extension to a granular gas.

\subsection{Normal gas}

The Boltzmann equation for the one particle reduced distribution function $f(%
\mathbf{r},\mathbf{v},t)$ is
\begin{equation}
\left( \partial _{t}+\mathbf{v\cdot }\boldsymbol{\nabla }\right) f=\mathcal{C%
}[f,f].  \label{3.0}
\end{equation}%
where $\mathcal{C}[f,f]$ describes uncorrelated elastic binary collisions
\cite{McL89}. The notation denotes that $\mathcal{C}[f,f]$ is a bilinear
functional of $f$. Its detailed form will not be required here, only the
fact that particle number, energy, and momentum are conserved as represented
by the properties%
\begin{equation}
\int d\mathbf{v}a_{\alpha }\mathcal{C}[f,f]=0,\hspace{0.2in}a_{\alpha }(%
\mathbf{v})\leftrightarrow \left( 1,\frac{1}{2}mv^{2},m\mathbf{v}\right) .
\label{3.1}
\end{equation}%
The set of functions $\left\{ a_{\alpha }\right\} $ are known as the
summational invariants. Their averages are also the hydrodynamic fields of
the last section
\begin{equation}
y_{\alpha }(\mathbf{r},t)=\int d\mathbf{v}a_{\alpha }\left( \mathbf{v}%
\right) f(\mathbf{r},\mathbf{v},t).  \label{3.2}
\end{equation}%
The macroscopic balance equations (\ref{2.1})-(\ref{2.3}), or in the
equivalent compact form (\ref{2.9}), follow from this definition by
differentiation with respect to time and application of the Boltzmann
equation. In this way the fluxes are identified as
\begin{equation}
\mathbf{j}_{\alpha }(\mathbf{r},t)=\int d\mathbf{vb}_{\alpha }\left( \mathbf{%
v}\right) f(\mathbf{r},\mathbf{v},t),\hspace{0.2in}\mathbf{b}_{\alpha }(%
\mathbf{v})=\mathbf{v}a_{\alpha }\left( \mathbf{v}\right) .  \label{3.3}
\end{equation}

Hydrodynamic equations result from these exact consequences of Boltzmann's
equation when the solution $f$ approaches a "normal" form on some length and
time scale, expected to be long compared to the mean free space and time
\cite{Dufty03}. A normal distribution is one for which all space and time
dependence occurs through the hydrodynamic fields%
\begin{equation}
f(\mathbf{r},\mathbf{v},t)\rightarrow f(\mathbf{v\mid }\left\{ y_{\alpha
}\right\} ).  \label{3.4}
\end{equation}%
The notation $f(\mathbf{v\mid }\left\{ y_{\alpha }\right\} )$ indicates a
\textit{functional} of the fields $y_{\alpha }(\mathbf{r},t)$ throughout the
system (equivalently, and for practical purposes, it is a \textit{function}
of the fields and all their derivatives at the point of interest). An
example of a normal distribution is the local equilibrium distribution
\begin{equation}
f_{\ell }(V)\equiv n\left( \frac{m}{2\pi T}\right) ^{3/2}\exp \left(
-mV^{2}/2T\right) .  \label{3.5}
\end{equation}%
where $\mathbf{V}\equiv \mathbf{v}-\mathbf{u}(\mathbf{r},t)$ and in all of
the following, units are used such that Boltzmann's constant $k_{B}=1$. This
distribution is parameterized by five fields $n$, $T$, and $\mathbf{u}$
which, as the notation suggests, are chosen to be the \emph{same} as the
hydrodynamic fields defined in (\ref{3.2}) and (\ref{2.7}). This is
expressed by the condition%
\begin{equation}
\int d\mathbf{v}a_{\alpha }\left( f-f_{\ell }\right) =0.  \label{3.6}
\end{equation}%
The local equilibrium distribution function depends on the fields, but not
their gradients. In fact, it is a solution to the Boltzmann equation to
zeroth order in the gradients, as follows from the second important property
of the collision operator%
\begin{equation}
\mathcal{C}[f_{\ell },f_{\ell }]=0.  \label{3.7}
\end{equation}%
More generally, the normal solution to the Boltzmann equation can be
constructed as an expansion in the gradients with $f_{\ell }$ as the leading
order contribution. This is done by the familiar Chapman-Enskog procedure
\cite{McL89} and is carried out in the Appendix.

It is useful to provide a geometrical representation of the normal solution.
First, define a set of functions "conjugate" to the $\left\{ a_{\alpha
}\right\} $ by

\begin{equation}
\psi _{\nu }=\frac{\partial f_{\ell }}{\partial y_{\nu }},\hspace{0.2in}\int
d\mathbf{v}a_{\alpha }\psi _{\nu }=\delta _{\alpha \nu }.  \label{3.8}
\end{equation}%
The second equality follows from (\ref{3.2}) and (\ref{3.6}) and shows the
sense in which $\left\{ a_{\alpha }\right\} $ and $\left\{ \psi _{\alpha
}\right\} $ form a biorthogonal set. Next, define the projection operator
whose action on an arbitrary distribution function $h$ is%
\begin{equation}
\mathcal{P}h=\psi _{\nu }\int d\mathbf{v}a_{\nu }h.  \label{3.9}
\end{equation}%
It is easily verified that $\mathcal{P}$ has the property of a projection
operator $\mathcal{P}^{2}=\mathcal{P}$. The utility of these definitions is
the decomposition of $f$ into its local equilibrium distribution plus a
remainder that is in the orthogonal subspace defined by $\mathcal{Q}=1-%
\mathcal{P}$
\begin{equation}
f=f_{\ell }+\delta f=f_{\ell }+\mathcal{P}\delta f+\mathcal{Q}\delta
f=f_{\ell }+\mathcal{Q}\delta f.  \label{3.10}
\end{equation}%
The last equality follows from (\ref{3.6}), $\mathcal{P}\delta f=0$. Since $%
f_{\ell }$ is a function of the fields, and not their gradients, all
contributions to a normal solution due to gradients must come from $\delta f=%
\mathcal{Q}\delta f$. Furthermore, since $f_{\ell }$ is a solution to the
Boltzmann equation at zeroth order in the gradients (see (\ref{3.7})), $%
\delta f=\mathcal{Q}\delta f$ is at least of first order. This allows a
decomposition of the fluxes (\ref{3.3}) into Euler and irreversible
contributions as in the previous section with the identifications
\begin{equation}
\mathbf{j}_{\alpha }^{(0)}(\mathbf{r},t)=\int d\mathbf{vb}_{\alpha }f_{\ell
},\hspace{0.2in}\mathbf{j}_{\alpha }^{\ast }(\mathbf{r},t)=\int d\mathbf{vb}%
_{\alpha }\mathcal{Q}\delta f\mathbf{.}  \label{3.11}
\end{equation}%
An immediate consequence of the property $\mathbf{b}_{1}\left( \mathbf{V}%
\right) =\mathbf{a}_{3}\left( \mathbf{V}\right) $ is $\mathbf{j}_{1}^{\ast
}=0$, resulting in the continuity equation for the number density.

The Boltzmann equation determines the detailed form for $\delta f$. Assuming
a normal form, the Boltzmann equation to first order in the gradients
becomes (see Appendix)%
\begin{equation}
L\delta f=-\left( \mathcal{Q}\boldsymbol{\gamma }_{\alpha }\right) \mathbf{%
\cdot \nabla }_{\mathbf{r}}\chi _{\alpha },  \label{3.12}
\end{equation}%
where $L$ is the linear Boltzmann collision operator%
\begin{equation}
Lh\equiv -\mathcal{C}[f_{\ell },h]-\mathcal{C}[h,f_{\ell }].  \label{3.13}
\end{equation}%
A transformation to the conjugate variables has been made using%
\begin{equation}
\mathbf{\nabla }_{\mathbf{r}}y_{\alpha }=g_{\alpha \nu }^{-1}\mathbf{\nabla }%
_{\mathbf{r}}\chi _{\nu },\hspace{0.3in}g_{\nu \alpha }=g_{\alpha \nu }=%
\frac{\partial \chi _{\alpha }}{\partial y_{\nu }},  \label{3.14a}
\end{equation}%
and $\mathbf{\gamma }_{\alpha }(\mathbf{v})$ are the conjugate fluxes are
defined by%
\begin{equation}
\mathbf{\gamma }_{\alpha }(\mathbf{v})=\mathbf{v}\psi _{\nu }\left( \mathbf{v%
}\right) g_{\nu \alpha }^{-1}=\mathbf{v}\frac{\partial f_{\ell }(\mathbf{v})%
}{\partial \chi _{\alpha }}  \label{3.14}
\end{equation}%
The appearence of the orthogonal projection $\mathcal{Q}$ on the right side
of (\ref{3.12}) assures that solutions to this equation exist. This is given
by the Fredholm alternative for such linear inhomogeneous equations which
states that the right side must be orthogonal to the null space for the
adjoint of $\mathcal{L}$, in this case given by $\mathcal{PL}=0$. The formal
solution can be written%
\begin{equation}
\delta f=-\mathcal{Q}\int_{0}^{\infty }dte^{-Lt}\mathcal{Q}\mathbf{\gamma }%
_{\alpha }(\mathbf{V})\mathbf{\cdot \nabla }_{\mathbf{r}}\chi _{\alpha },
\label{3.15}
\end{equation}%
In general, an arbitrary solution to the homogeneous equation $L\delta f=0$
could be added to this, but the condition (\ref{3.6}), $\mathcal{P\delta }%
f=0 $ excludes such terms.

Use of this formal solution in (\ref{3.11}) gives the linear constitutive
equations (\ref{2.11}), and the transport matrix $L_{\alpha \beta }$ is
identified as%
\begin{equation}
\mathrm{L}_{\alpha \beta }=\int_{0}^{\infty }d\tau \mathrm{C}_{\alpha \beta
}(\tau ),  \label{3.17}
\end{equation}%
\begin{equation}
\mathrm{C}_{\alpha \beta }(\tau )=\int d\mathbf{v}\left( \mathcal{Q}%
^{\dagger }\mathbf{b}_{\alpha }\right) e^{-L\tau }\mathcal{Q}\mathbf{\gamma }%
_{\beta }.  \label{3.18}
\end{equation}%
where $\mathcal{Q}^{\dagger }$ is the adjoint of $\mathcal{Q}$. This makes
explicit the low density form for the general phenomenological postulate. It
is given in the form of a Green-Kubo expression, where the "time correlation
function" $\mathrm{C}_{\alpha \beta }(\tau )$ is a flux - conjugate flux
correlation function. Further simplifications are possible by noting that
both the conjugate densities $\psi _{\nu }$ and conjugate fluxes $\mathbf{%
\gamma }_{\nu }$ can be written as linear cominations of the densities $%
a_{\nu }$ and $\mathbf{b}_{\nu }$, respectively
\begin{equation}
\psi _{\nu }=f_{\ell }\sum_{\sigma }c_{\nu \sigma }a_{\alpha },\quad \mathbf{%
\gamma }_{\nu }=f_{\ell }\sum_{\sigma }c_{\nu \sigma }\mathbf{b}_{\sigma }.
\label{3.19}
\end{equation}%
\newline
Note that the (adjoint) projection operator $\mathcal{Q}^{\dagger }$ implies
that $\mathrm{C}_{1\beta }(\tau )=0$ since $\mathbf{b}_{1}\propto \mathbf{a}%
_{3}$ (the mass flux is the momentum density) and $\mathcal{Q}^{\dagger }$
projects orthogonal to the set of densities $\left\{ a_{\alpha }\right\} $.

The $\mu $ coefficient vanishes for similar reasons. It is given by
\begin{equation}
\mu =\int_{0}^{\infty }d\tau C_{21}(\tau )\frac{\partial \chi _{1}}{\partial
n}\mid _{T}.  \label{3.20}
\end{equation}%
The conjugate flux $\mathbf{\gamma }_{1}$ is proportional to the conjugate
density $\boldsymbol{\psi }_{3}$%
\begin{equation}
\mathbf{\gamma }_{1}=\mathbf{v}\frac{\partial f_{\ell }}{\partial \chi _{1}}%
\mid _{T}=-\frac{T}{mn}\frac{\partial \chi _{1}}{\partial n}\mid _{T}%
\boldsymbol{\psi }_{3}.  \label{3.22}
\end{equation}%
Since $\mathcal{Q}$ projects orthogonal to the conjugate densities $\left\{
\psi _{\alpha }\right\} $, i.e. $\mathcal{Q}\boldsymbol{\psi }_{3}=0$, the
projected flux $\mathcal{Q}\mathbf{\gamma }_{\beta }$, the correlation
function $C_{21}(t)$, and consequently $\mu $ all vanish
\begin{equation}
\mu =-\frac{T}{mn}\frac{\partial \chi _{1}}{\partial n}\mid
_{T}\int_{0}^{\infty }d\tau \int d\mathbf{v}\left( \mathcal{Q}^{\dagger }%
\mathbf{b}_{\alpha }\right) _{2}e^{-L\tau }\mathcal{Q}\boldsymbol{\psi }%
_{3}=0,  \label{3.23}
\end{equation}

In summary, the correlation functions determining the transport coefficients
involve fluxes projected orthogonal to associated densities. The fluxes
associated with the density are themselves densities, and hence such
correlation functions vanish. In particular those associated with
dissipation in the continuity equation and the contribution to the heat flux
from density gradients are zero.

\subsection{Granular gas}

The analysis for the granular Boltzmann equation proceeds in a similar way
\cite{BDKyS98}, and the details are given in Appendix A. The correlation
functions for the transport matrix $\mathrm{L}_{\alpha \beta }$ in this more
general case are%
\begin{equation}
\mathrm{C}_{\alpha \beta }(\tau )=\int d\mathbf{v}\left( \mathcal{Q}^{\dag }%
\mathbf{b}_{\alpha }\right) e^{-\mathcal{L}\tau }\left( e^{\Lambda \tau }%
\mathcal{Q}\boldsymbol{\gamma }\right) _{\beta }.  \label{3.24}
\end{equation}%
\begin{equation}
\mathcal{L}\equiv \frac{1}{2}\xi \boldsymbol{\nabla }_{\mathbf{V}}\cdot
\mathbf{V}+L,\hspace{0.3in}\xi \equiv -\frac{2}{dnT}w\left[ f^{(0)},f^{(0)}%
\right] .  \label{3.24a}
\end{equation}%
The parameter $\xi $ is called the cooling rate since it determines the
relative rate of change of the temperature in the HCS due to inelastic
collisions. The densities $\left\{ a_{\alpha }\right\} $\textbf{\ }and
fluxes $\left\{ \mathbf{b}_{\alpha }\right\} $ are the same as those in (\ref%
{3.1}) and (\ref{3.3}). The projection operator is given by (\ref{3.3}) but
with new conjugate densities $\left\{ \psi _{\alpha }\right\} $\textbf{\ }%
and fluxes $\left\{ \boldsymbol{\gamma }_{\alpha }\right\} $ defined in
terms of the reference local homogeneous cooling state (HCS) $f^{(0)}$
instead of the local equilibrium state $f_{\ell }\hspace{0.2in}$%
\begin{equation}
\psi _{\nu }=\frac{\partial f^{(0)}}{\partial y_{\nu }},\hspace{0.2in}%
\mathbf{\gamma }_{\alpha }=\mathbf{v}\frac{\partial f^{(0)}(\mathbf{v})}{%
\partial \chi _{\alpha }}.  \label{3.25}
\end{equation}%
The reference HCS is the normal solution to the Boltzmann equation in the
absence of gradients%
\begin{equation}
\frac{1}{2}\xi \boldsymbol{\nabla }_{\mathbf{V}}\cdot \left( \mathbf{V}%
f^{(0)}\right) =\mathcal{C}[f^{(0)},f^{(0)}].  \label{3.26}
\end{equation}%
The cooling rate $\xi $ vanishes in the elastic limit and $%
f^{(0)}\rightarrow f_{\ell }$ in this limit.

Aside from these differences in the reference state for granular fluids, the
generator for dynamics in (\ref{3.24}) shows significant differences. The
linearized operator $L$ is now that associated with the Boltzmann collision
operator for inelastic collisions. In addition, there is a velocity scaling
operator $\frac{1}{2}\xi \boldsymbol{\nabla }_{\mathbf{V}}\cdot \mathbf{V}$
(the derivative operates on everything to its right) which compensates for
the cooling generated by the collisions, as illustrated in the exact balance
of these effects in (\ref{3.26}). Finally, there is the matrix $\Lambda $
whose explicit form is given in (\ref{a.18}) and (\ref{a.20}) of the
Appendix. The relevant point here is that its eigenvalues $\left( 0,\frac{1}{%
2}\xi ,-\frac{1}{2}\xi \right) $ are the same as the smallest eigenvalues of
$\mathcal{L}$. Thus, the entire generator for the dynamics has a null space,
just as for normal fluids, and the projection $\mathcal{Q}$ assures that the
fluxes are orthogonal to the invariants defining this null space. This is a
necessary condition for the existence of the integral defining the transport
matrix $\mathrm{L}_{\alpha \beta }$.

It is seen that the condition for the continuity equation still holds $%
\mathrm{C}_{1\beta }(\tau )=0$, for the same reason as in normal fluids, $%
\mathcal{Q}^{\dag }\mathbf{b}_{1}\propto \mathcal{Q}^{\dag }\mathbf{a}_{3}=0$%
. However, the related conditions for the transport coefficient $\mu $ to
vanish are no longer satisfied in general. The relevant correlation function
$\mathrm{C}_{21}(\tau )$ is no longer determined by the single flux $%
\boldsymbol{\gamma }_{1}$ but is coupled to $\boldsymbol{\gamma }_{2}$ as
well. Neither of these fluxes in simply proportional to the conjugate
densities $\left\{ \psi _{\alpha }\right\} $ and hence the action of $%
\mathcal{Q}$ on each is non zero. This difference occurs for granular fluids
because $f^{(0)}$ is no longer determined entirely in terms of the $\left\{
a_{\alpha }\right\} $ and hence there is no simple relationship between the
sets $\left\{ a_{\alpha }\right\} $\textbf{,}$\left\{ \mathbf{b}_{\alpha
}\right\} $ and $\left\{ \psi _{\alpha }\right\} $\textbf{,}$\left\{
\boldsymbol{\gamma }_{\alpha }\right\} $ as expressed in (\ref{3.19}). The
explicit form for $\mu $ is obtained in Appendix A%
\begin{eqnarray}
\mu  &=&-\int_{0}^{\infty }d\tau \int d\mathbf{v}\left( \mathcal{Q}^{\dag }%
\mathbf{b}_{2}\right) \left\{ e^{-\left( \mathcal{L}-\lambda _{1}\right)
\tau }\mathcal{Q}\mathbf{v}\frac{\partial f^{(0)}}{\partial n}\mid
_{T}\right.   \notag \\
&&\left. -2\frac{\partial e}{\partial n}\mid _{T}\left( e^{-\left( \mathcal{L%
}-\lambda _{2}\right) \tau }-e^{-\left( \mathcal{L}-\lambda _{1}\right) \tau
}\right) \mathcal{Q}\left( \mathbf{v}\frac{\partial f^{(0)}}{\partial e}\mid
_{n}\right) \right\} .  \label{3.27}
\end{eqnarray}%
The new conjugate densities $\left\{ \psi _{\alpha }\right\} $ are
invariants of the new dynamics generated by $\left( \mathcal{L}-\lambda
_{\alpha }\right) $. The projection operators again assure that there is no
contribution from the invariants which is a necessary condition for
convergence of the $\tau $ integral. In general, however, neither $\mathbf{v}%
\partial f^{(0)}/\partial n\mid _{T}$ nor $\mathbf{v}\partial
f^{(0)}/\partial e\mid _{T}$is a linear combination of these invariants and
the action of $\mathcal{Q}$ on them does not vanish. Only in the elastic
limit does $\mathcal{Q}\mathbf{v}\partial f^{(0)}/\partial n\mid _{T}$
become proportional to $\mathcal{Q}\boldsymbol{\psi }_{3}=0$. In this limit $%
\lambda _{2}=\lambda _{1}=0$ and the coefficient of $\mathbf{v}\partial
f^{(0)}/\partial e\mid _{T}$ also vanishes, confirming $\mu =0$ for a normal
fluid.

\section{Formal Linear Response}

\label{sec5}The analysis of hydrodynamics from kinetic theory can be
generalized by the formal application of nonequilibrium statistical
mechanics to granular fluids. The details are described in references \cite%
{DBB07,BDB07}. The starting point is the Liouville equation for the $N$
particle phase space density $\rho \left( \Gamma ,t\right) $, where $\Gamma
\equiv \{\mathbf{q}_{1},..\mathbf{q}_{N},\mathbf{v}_{1},..\mathbf{v}_{N}\}$
denotes a point in the $6N$ dimensional phase space \cite{Brey97}. First, a
homogeneous normal solution to the Liouville equation is $\rho _{h}\left(
\Gamma \right) $ identified, representing the homogeneous cooling state
(HCS) for an isolated system%
\begin{equation}
\overline{\mathcal{L}}\rho _{h}=0,\hspace{0.3in}\overline{\mathcal{L}}%
X\equiv \frac{1}{2}\zeta _{h}\sum_{i=1}^{N}\nabla _{\mathbf{V}_{i}}\cdot (%
\mathbf{V}_{i}X)+\overline{L}X.  \label{4.1}
\end{equation}%
Here $\overline{L}$ is the Liouville operator for $N$ hard inelastic spheres
and $\zeta _{h}$ is the associated cooling rate in the HCS. In addition,
there is a scaling operator $\frac{1}{2}\zeta _{h}\nabla _{\mathbf{V}%
_{i}}\cdot \mathbf{V}_{i}$ for each particle. Clearly, $\overline{\mathcal{L}%
}$ is the $N$ particle generalization of the kinetic theory generator $%
\mathcal{L}$ of (\ref{3.24a}). Next, small spatial perturbations of this
state are induced through an associated local HCS $\rho _{\ell h}$. The
response of the hydrodynamic fields at a later time due to these initial
perturbations is characterized by response functions, represented as time
correlation functions composed of the $N$ particle phase functions for the
hydrodynamic fields (corresponding to the $a_{\alpha }$ of the last section)
and the functional derivatives of $\rho _{\ell h}$ with respect to the
conjugate fields $\left\{ \chi _{\alpha }\right\} $ (corresponding to the $%
\psi _{\alpha }$ of the last section). Since these response functions must
exhibit hydrodynamic excitations at long times and long wavelengths, the
transport matrix $\mathrm{L}_{\alpha \beta }$ can be identified. It has the
representation in terms of time correlation functions again, as in (\ref%
{3.17}) but now the results are formally exact without the restrictions of
the kinetic theory.

For the purposes here it is sufficient to display only the final result for
the coefficient $\mu $. It has a form similar to that of (\ref{3.20})%
\begin{equation}
\mu =\mu _{0}+\int_{0}^{\infty }d\tau C_{21}(\tau )\frac{\partial \chi _{1}}{%
\partial n}\mid _{T}.  \label{4.2}
\end{equation}%
There is an additional term $\mu _{0}$ that does not have the form of a time
integral of a correlation function, and is due to both the singular hard
sphere dynamics and the dissipation. It vanishes in the low density limit,
and so does not appear at the level of the Boltzmann equation. It is given by%
\begin{equation}
\mu _{0}=\frac{e}{dT}V^{-1}\int d\Gamma \,\mathbf{S}(\Gamma )\cdot \mathbf{M}%
(\Gamma ).  \label{4.3}
\end{equation}%
where$\,\mathbf{S}(\Gamma )$ is the volume integrated phase function for the
energy flux (corresponding to $\mathbf{b}_{2}$ in the kinetic theory
analysis), and $\mathbf{M}(\Gamma )$ is the space moment for the functional
derivative of $\rho _{lh}$ with respect to density%
\begin{equation}
\mathbf{M}=\int d\mathbf{rr}\left( \frac{\delta \rho _{lh}}{\delta n\left(
\mathbf{r}\right) }|_{T}\right) _{\delta y=0}.  \label{4.4}
\end{equation}%
It is easily seen that $\mu _{0}\rightarrow 0$ in the elastic limit, for
which $\rho _{lh}$ becomes the corresponding equilibrium ensemble. Otherwise
it is non-zero. The correlation function in (\ref{4.2}) has a form similar
to that of (\ref{3.24})%
\begin{equation}
C_{21}(\tau )\frac{\partial \chi _{1}}{\partial n}\mid _{T}=V^{-1}\int
d\Gamma \,\left( \mathcal{Q}^{\dag }\mathbf{S}\right) e^{-\overline{\mathcal{%
L}}\tau }\left( e^{\Lambda \tau }\mathcal{Q}\boldsymbol{\Upsilon }\right)
_{1}  \label{4.5}
\end{equation}%
where the conjugate flux is%
\begin{equation}
\boldsymbol{\Upsilon }_{\alpha }\equiv -\left( \overline{\mathcal{L}}%
-\Lambda \right) \mathbf{M}_{\alpha }  \label{4.6}
\end{equation}%
Here $\mathbf{M}_{\alpha }$ is the space moment of the functional derivative
of $\rho _{lh}$ with respect to $y_{\alpha }$ holding the other $\left\{
y_{\beta }\right\} $ constant, and evaluated at the HCS.

The projection operator $\mathcal{Q}$ projects orthogonal to the invariants
of the dynamics generated by $\left( \overline{\mathcal{L}}-\Lambda \right) $%
. Consider now the elastic limit for which $\rho _{lh}$ becomes a local
equilibrium ensemble $\rho _{le}$. To be specific, let that be the local
grand canonical ensemble. The correlation function simplifies to%
\begin{equation}
C_{21}(\tau )\frac{\partial \chi _{1}}{\partial n}|_{T}\rightarrow
-V^{-1}\int d\Gamma \,\left( \mathcal{Q}^{\dag }\mathbf{S}\right) e^{-%
\overline{L}\tau }\mathcal{Q}\overline{L}\mathbf{M,}  \label{4.6a}
\end{equation}%
\begin{equation}
\mathbf{M}\rightarrow -\rho _{e}m\int d\mathbf{rr}\int d\mathbf{r}^{\prime
}\left( \frac{\delta \chi _{1}\left( \mathbf{r}^{\prime }\right) }{\delta
n\left( \mathbf{r}\right) }|_{T}\right) _{\delta y=0}\widehat{n}\left(
\mathbf{r}^{\prime }\right) ,  \label{4.7}
\end{equation}%
where $\nu $ is the activity, $\rho _{e}$ is the strict equilibrium
ensemble, and $\widehat{n}\left( \mathbf{r}\right) $ is the phase function
corresponding to the number density. Then using $\overline{L}\widehat{n}%
\left( \mathbf{r}\right) =-m^{-1}\nabla \cdot \widehat{\mathbf{g}}\left(
\mathbf{r}\right) ,$ where $\widehat{\mathbf{g}}\left( \mathbf{r}\right) $
is the phase function representing the momentum density, the flux $\overline{%
L}\mathbf{M}$ is found to be proportional to the total momentum $\mathbf{P}$
\begin{eqnarray}
\overline{L}\mathbf{M} &\mathbf{=}&\rho _{e}m\int d\mathbf{rr}\int d\mathbf{r%
}^{\prime }\left( \frac{\delta \chi _{1}\left( \mathbf{r}^{\prime }\right) }{%
\delta n\left( \mathbf{r}\right) }|_{T}\right) _{\delta y=0}\overline{L}%
n\left( \mathbf{r}^{\prime }\right)  \notag \\
&=&-\rho _{e}\int d\mathbf{rr}\int d\mathbf{r}^{\prime }\left( \frac{\delta
\chi _{1}\left( \mathbf{r}^{\prime }\right) }{\delta n\left( \mathbf{r}%
\right) }|_{T}\right) _{\delta y=0}\nabla ^{\prime }\cdot g\left( \mathbf{r}%
^{\prime }\right)  \notag \\
&=&\rho _{e}\frac{\partial \chi _{1}}{\partial n}|_{T}\mathbf{P}.
\label{4.8}
\end{eqnarray}%
Consequently, $\mathcal{Q}\overline{L}\mathbf{M=0}$ and $C_{21}(\tau )=0$.
It was already noted that $\mu _{0}$ vanishes in the elastic limit, so $\mu
=0$ as well and the usual form of Fourier's law is recovered.

More generally, for inelastic collisions $\mu _{0}\neq 0$ (except at low
density) and none of the fluxes $\boldsymbol{\Upsilon }$ are simply
proportional to the invariants. Consequently, $\mathcal{Q}\boldsymbol{%
\Upsilon }\neq 0$ and $\mu \neq 0$.

\section{Discussion}

The theoretical "discovery" that Fourier's law for a granular fluid has an
additional term proportional to the density gradient provides an interesting
qualitative difference from normal fluids. There has been much discussion
about this additional term and attempts to detect it in simulations or
experiments. The theoretical analysis here provides a different perspective,
in hindsight, that the surprising nature of this difference between
Navier-Stokes hydrodynamics for normal and granular fluids is the \textit{%
absence} of this term in the former case rather than its presence for the
latter case. This circumstance is similar to the discovery of generic long
range correlations in nonequilibrium states, absent at equilibrium \cite%
{Sengers06}. Both provide examples of the very special balance of competing
effects for the equilibrium state, in contrast to the qualitatively
different behavior for generic nonequilibrium states.

The vanishing of the transport coefficient for the contribution to the heat
flux from a density gradient follows from the simple structure of the local
conservation law for the number density, and the characterization of the
equilibrium state in terms of the dynamical invariants. The number
conservation law relates the density to its flux. However, this flux is
itself proportional to a density for one of the invariants - the momentum.
The transport coefficients are time integrals of correlation functions
composed of these fluxes and corresponding "conjugate" fluxes generated from
the equilibrium ensemble. Convergence of the time integrals requires that
there be no time independent parts to these correlation functions. Thus, the
correlation functions are constructed from those parts of the fluxes that
have their invariant parts subtracted (or projected) out. As the number flux
is itself an invariant all transport processes coupling to the number flux
will therefore have zero transport coefficients. This leads directly to the
vanishing of all dissipative contributions to the continuity equation. In
addition, since the equilibrium ensemble is a function of the invariants it
results that the conjugate fluxes are linear combinations of the fluxes in
the conservation laws. Thus any contributions from the number flux, itself a
density, has no remainder once its invariant parts are subtracted out. This
leads to the vanishing of the coefficient $\mu $ in Fourier's law.

The simple structure of the conservation law for number density does not
depend on the state of the system considered. Thus, there are no dissipative
contributions to the number flux for any state, equilibrium or
nonequilibrium. In contrast, the conjugate fluxes depend in detail on the
reference state about which hydrodynamic excitations are being considered.
The simple relationship of the conjugate fluxes to the fluxes in the
conservation laws for equilibrium states cannot be expected more generally
for any nonequilibrium reference state. This is the case for both normal and
granular fluids. An example of the former is the hydrodynamic excitations
about uniform shear flow \cite{Lee97} where Fourier's law in the form of Eq.
(\ref{1.1}) applies with $\mu \neq 0$.

In closing, it may be useful to display concrete expressions for $\lambda $
and $\mu $ in a granular gas obtained from an approximate evaluation of the
expression given above from the Boltzmann equation \cite{ApproxMu} for $d=3$%
\begin{equation*}
\lambda =\frac{nT}{m}\frac{5}{2\left( \nu -2\xi \right) },\hspace{0.3in}\mu =%
\frac{T}{n}\lambda \frac{2\xi }{2\nu -3\xi }
\end{equation*}%
where $\xi =5\nu _{0}\left( 1-\alpha ^{2}\right) /12$ is the cooling rate of
(\ref{3.26}) calculated in this same approximation, $\nu =\nu _{0}\left(
1+\alpha \right) (49-33\alpha )/48$, and $\nu _{0}=16n\sigma ^{2}\left( \pi
T/m\right) ^{1/2}/5$ is an average collision rate. In the nearly elastic
limit $\lambda \rightarrow 15nT/4m\nu _{0}$ and $\mu \rightarrow \left(
1-\alpha \right) 75T^{2}/16m\nu _{0}.$ The simulations of references \cite%
{Brey04} and \cite{Brey05} confirm the more general theoretical prediction
over a wide range of values for the restitution coefficient $\alpha $.

\section{Acknowledgements}

Comments from A. Baskaran, Syracuse University, J. J. Brey, Universidad de
Sevilla, and A. Santos, Universidad de Extremadura, are gratefully
acknowledged.

\appendix

\section{Normal Solution to Boltzmann Equation}

\label{ap1} In this Appendix the normal solution to the Boltzmann equation
for a granular gas is obtained up through first order in the gradients \cite%
{BDKyS98},
\begin{equation}
f=f^{(0)}+\mathbf{F}_{\alpha }\cdot \boldsymbol{\nabla }_{\mathbf{r}%
}y_{\alpha }+\cdot \cdot  \label{a.1}
\end{equation}%
The leading order term $f^{(0)}$ and coefficient $\mathbf{F}_{\alpha }$ are
functions of the actual hydrodynamic fields $\left\{ y_{a}\right\} $. As a
normal solution obeys a condition analogous to (\ref{3.6})
\begin{equation}
\int d\mathbf{v}a_{\alpha }\left( f-f^{(0)}\right) =0.  \label{a.1.1}
\end{equation}%
Only the case of hard sphere interactions is considered, so there is no
internal energy scale. Then, from dimensional analysis, they have the forms
\begin{equation}
f^{(0)}\equiv n\left( \frac{m}{2\pi T}\right) ^{3/2}f^{(0)\ast }\left(
V^{\ast }\right) ,\hspace{0.3in},\hspace{0.3in}\mathbf{V}^{\ast }=\sqrt{%
\frac{m}{2T}}\left( \mathbf{v}-\mathbf{U}\right) \mathbf{,}  \label{a.2}
\end{equation}%
\begin{equation}
\mathbf{F}_{1}=\left( \frac{m}{2T}\right) ^{3/2}\mathbf{F}_{1}^{\ast }(%
\mathbf{V}^{\ast }),\hspace{0.3in}\mathbf{F}_{2}=\frac{2}{dT}\left( \frac{m}{%
2T}\right) ^{3/2}\mathbf{F}_{\alpha }^{\ast }(\mathbf{V}^{\ast }),\hspace{%
0.3in}\mathbf{F}_{3}=\frac{1}{m}\left( \frac{m}{2T}\right) ^{2}\mathbf{F}%
_{\alpha }^{\ast }(\mathbf{V}^{\ast }).  \label{a.3}
\end{equation}%
The temperature $T$ for hard spheres is related to the energy by $e=\frac{3}{%
2}nT+\frac{1}{2}mnU^{2}$, and the asterisk denotes a dimensionless quantity.
Substitution of (\ref{a.1}) into the Boltzmann equation gives
\begin{equation}
\left( \frac{\partial f^{(0)}}{\partial y_{\sigma }}+\frac{\partial \mathbf{F%
}_{\alpha }}{\partial y_{\sigma }}\cdot \boldsymbol{\nabla }_{\mathbf{r}%
}y_{\alpha }+\mathbf{F}_{\sigma }\cdot \mathbf{\nabla }+\cdot \cdot \right)
\left( \partial _{t}y_{\sigma }+\mathbf{v\cdot }\boldsymbol{\nabla }_{%
\mathbf{r}}y_{\sigma }\right) =\mathcal{C}[f,f],  \label{a.4}
\end{equation}%
and using the macroscopic balance equations (\ref{2.9}) the time derivative
can be expressed in terms of the gradients
\begin{equation}
\left( \frac{\partial f^{(0)}}{\partial y_{\sigma }}+\frac{\partial \mathbf{F%
}_{\alpha }}{\partial y_{\sigma }}\cdot \boldsymbol{\nabla }_{\mathbf{r}%
}y_{\alpha }+\mathbf{F}_{\sigma }\cdot \mathbf{\nabla }+\cdot \cdot \right)
\left( w_{\sigma }-\boldsymbol{\nabla }\cdot \mathbf{j}_{\sigma }+\mathbf{%
v\cdot }\boldsymbol{\nabla }_{\mathbf{r}}y_{\sigma }\right) =\mathcal{C}%
[f,f].  \label{a.5}
\end{equation}

To zeroth order in the gradients this equation determines $f^{(0)}$
\begin{equation}
\frac{\partial f^{(0)}}{\partial y_{\alpha }}w_{\alpha }\left[
f^{(0)},f^{(0)}\right] =\mathcal{C}[f^{(0)},f^{(0)}],  \label{a.6}
\end{equation}%
where the bilinear functional dependence of the energy loss $w_{\sigma }$
has been made explicit to show that here it is evaluated to lowest order.
Since the source $w_{\alpha }$ occurs only in the energy equation the left
side of this equation can be made more explicit
\begin{equation}
\frac{\partial f^{(0)}}{\partial y_{\sigma }}w_{\sigma }\left[
f^{(0)},f^{(0)}\right] =\frac{\partial f^{(0)}}{\partial e}w\left[ f^{(0)}%
\right] =\frac{\partial f^{(0)}}{\partial T}\frac{2}{3n}w\left[
f^{(0)},f^{(0)}\right] =\frac{1}{2}\xi \left( d+\mathbf{V}\cdot \mathbf{%
\nabla }\right) f^{(0)},  \label{a.7}
\end{equation}%
and the cooling rate $\xi \equiv -2w\left[ f^{(0)},f^{(0)}\right] /3nT$ has
been introduced. Equation (\ref{a.6}) for $f^{(0)}$ becomes
\begin{equation}
\frac{1}{2}\xi \boldsymbol{\nabla }_{\mathbf{V}}\cdot \left( \mathbf{V}%
f^{(0)}\right) =\mathcal{C}[f^{(0)},f^{(0)}].  \label{a.8}
\end{equation}%
Although simple analytic forms for the solution to this equation have not
yet been found, its behavior for small and large velocities is known, and
good approximations exist more generally. Furthermore, it has been studied
numerically using Direct Simulation Monte Carlo. For our purposes,
therefore, it can be considered as known. In the elastic limit the solution
to $\mathcal{C}[f^{(0)},f^{(0)}]=0$ is the Maxwellian (\ref{3.5}).

To first order in the gradients Eq. (\ref{a.5}) determines $\mathbf{F}%
_{\alpha }$ as the solution to
\begin{equation}
\left[ L\mathbf{F}_{\alpha }+\left( w\frac{\partial \mathbf{F}_{\alpha }}{%
\partial e}+\frac{\partial w}{\partial y_{\alpha }}\mathbf{F}_{2}\right) %
\right] \cdot \boldsymbol{\nabla }_{\mathbf{r}}y_{\alpha }=-\frac{\partial
f^{(0)}}{\partial y_{\alpha }}\left( \mathbf{v\cdot }\boldsymbol{\nabla }_{%
\mathbf{r}}y_{\alpha }-\boldsymbol{\nabla }_{\mathbf{r}}\cdot \mathbf{j}%
_{\alpha }^{(0)}\right)  \label{a.9}
\end{equation}%
\begin{equation}
Lh\equiv -\mathcal{C}[f_{\ell },h]-\mathcal{C}[h,f_{\ell }].  \label{a.9a}
\end{equation}%
It is understood here that $w=w_{\alpha }\left[ f^{(0)},f^{(0)}\right] $.
The Euler order flux $\mathbf{j}_{\alpha }^{(0)}$ is defined in terms of the
lowest order distribution as in (\ref{3.11})
\begin{equation}
\mathbf{j}_{\alpha }^{(0)}(\mathbf{r},t)=\int d\mathbf{vb}_{\alpha }\left(
\mathbf{v}\right) f^{(0)}(\mathbf{r},\mathbf{v},t).  \label{a.10}
\end{equation}%
It follows directly that the second term on the right side of (\ref{a.9})
can be written
\begin{equation}
\frac{\partial f^{(0)}}{\partial y_{\alpha }}\boldsymbol{\nabla }_{\mathbf{r}%
}\cdot \mathbf{j}_{\alpha }^{(0)}(\mathbf{r},t)=\psi _{\alpha }\int d\mathbf{%
va}_{\alpha }\left( \mathbf{v}\right) \mathbf{v}\psi _{\alpha }\left(
\mathbf{v}\right) \cdot \boldsymbol{\nabla }_{\mathbf{r}}y_{\alpha }=%
\mathcal{P}\left( \mathbf{v}\psi _{\alpha }\left( \mathbf{v}\right) \right)
\cdot \boldsymbol{\nabla }_{\mathbf{r}}y_{\alpha },  \label{a.11}
\end{equation}%
where $\psi _{\alpha }$ \ and the projection operator $\mathcal{P}$ are
defined \textit{mutatis mutandis} as in (\ref{3.8}) and (\ref{3.9})%
\begin{equation}
\psi _{\nu }=\frac{\partial f^{(0)}}{\partial y_{\nu }},\hspace{0.2in}\int d%
\mathbf{v}a_{\alpha }\left( \mathbf{v}\right) \psi _{\nu }\left( \mathbf{v}%
\right) =\delta _{\alpha \nu },\hspace{0.2in}\mathcal{P}g(\mathbf{v})=\psi
_{\nu }(\mathbf{v})\int d\mathbf{v}a_{\nu }\left( \mathbf{v}\right) g\left(
\mathbf{v}\right) .  \label{a.11a}
\end{equation}%
The gradients of $y_{\alpha }$ are arbitrary in (\ref{a.9}) so their
coefficients give the desired equations for $\mathbf{F}_{\alpha }$%
\begin{equation}
\left( w\frac{\partial }{\partial e}+\mathcal{L}\right) \mathbf{F}_{\alpha }+%
\frac{\partial w}{\partial y_{\alpha }}\mathbf{F}_{2}=-\mathcal{Q}\left(
\mathbf{v}\psi _{\alpha }\left( \mathbf{v}\right) \right) ,\hspace{0.3in}%
\mathcal{Q}=1-\mathcal{P}.  \label{a.12}
\end{equation}

This can be simplified further by noting that $w\propto (nT)^{3/2}n^{1/2}$
and using the scaling of (\ref{a.3})%
\begin{equation}
w\frac{\partial }{\partial e}=\frac{2w}{dn}\frac{\partial }{\partial T},%
\hspace{0.3in}\frac{\partial w}{\partial n}=\frac{1}{2}\frac{w}{n},\hspace{%
0.3in}\frac{\partial w}{\partial e}=\frac{2}{3n}\frac{\partial w}{\partial T}%
=\frac{w}{nT},  \label{a.13}
\end{equation}%
to get%
\begin{equation}
\left( \mathcal{L}-\lambda _{\alpha }\right) \mathbf{F}_{\alpha }-\delta
_{\alpha 1}\xi \frac{3T}{4}\mathbf{F}_{2}=-\mathcal{Q}\left( \mathbf{v}\psi
_{\alpha }\left( \mathbf{v}\right) \right) ,  \label{a.14}
\end{equation}%
\begin{equation}
\mathcal{L}\equiv \frac{1}{2}\xi \boldsymbol{\nabla }_{\mathbf{V}}\cdot
\mathbf{V}+L.  \label{a.14a}
\end{equation}%
The constants $\lambda _{\alpha }$%
\begin{equation}
\lambda _{\alpha }\Leftrightarrow \left( 0,\frac{1}{2}\xi ,-\frac{1}{2}\xi
\right)  \label{a.15}
\end{equation}%
are eigenvalues of the operator $\frac{1}{2}\xi \boldsymbol{\nabla }_{%
\mathbf{V}}\cdot \mathbf{V}+\mathcal{L}$%
\begin{equation}
\left( \mathcal{L}-\lambda _{\alpha }\right) \varphi _{\alpha }=0
\label{a.16}
\end{equation}%
The eigenfunctions $\varphi _{\alpha }$ are linear combinations of the set $%
\left\{ \psi _{\alpha }\right\} $. Since the operator $\mathcal{Q}$ projects
orthogonal to this null space the Fredholm alternative is satisfied for
these integral equations and their solutions exist.

The set of equations (\ref{a.14}) can be written in matrix form%
\begin{equation}
\left( \mathcal{L}I-\overline{\Lambda }\right) _{\alpha \beta }\mathbf{F}%
_{\beta }=-\mathcal{Q}\left( \mathbf{v}\psi _{\alpha }\left( \mathbf{v}%
\right) \right) =-g_{\alpha \beta }^{-1}\mathcal{Q}\left( \boldsymbol{\gamma
}_{\beta }\left( \mathbf{v}\right) \right) ,  \label{a.17}
\end{equation}%
with%
\begin{equation}
\overline{\Lambda }_{\alpha \beta }=\lambda _{\alpha }\delta _{\alpha \beta
}-\frac{w}{2n}\delta _{\alpha 1}\delta _{\beta 2},  \label{a.18}
\end{equation}%
where $I$ is the identity matrix. Also, the fluxes $\boldsymbol{\gamma }%
_{\beta }$ are defined as in (\ref{3.14})%
\begin{equation}
\mathbf{\gamma }_{\alpha }(\mathbf{v})=\mathbf{v}\left( \psi g\right)
_{\alpha }=\mathbf{v}\frac{\partial f^{(0)}(\mathbf{v})}{\partial \chi
_{\alpha }}.  \label{a.19}
\end{equation}%
Next, perform a simlarity transformation with the symmetric matrix $g$ to
get the form%
\begin{equation}
\left( \mathcal{L}I-\Lambda \right) _{\alpha \beta }\left( g\mathbf{F}%
\right) _{\beta }=-\mathcal{Q}\left( \gamma _{\alpha }\left( \mathbf{v}%
\right) \right) ,\hspace{0.3in}\Lambda =g\overline{\Lambda }g^{-1}.
\label{a.20}
\end{equation}%
The solution (\ref{a.1}) can be given the representation%
\begin{equation}
\delta f=f-f^{(0)}=-\left[ \mathcal{Q}\int_{0}^{\infty }d\tau e^{-\mathcal{L}%
\tau }\left( e^{\Lambda \tau }\right) _{\alpha \beta }\mathcal{Q}\mathbf{%
\gamma }_{\beta }\right] \mathbf{\cdot \nabla }_{\mathbf{r}}\chi _{\alpha }.
\label{a.21}
\end{equation}%
The irreversible fluxes are obtained from (\ref{3.11})
\begin{equation}
\mathbf{j}_{\alpha }^{\ast }=\int d\mathbf{vb}_{\alpha }\mathcal{Q}\delta
f=\int d\mathbf{v}\left( \mathcal{Q}^{\dag }\mathbf{b}_{\alpha }\right) e^{-%
\mathcal{L}\tau }\left( e^{\Lambda s}\mathcal{Q}\boldsymbol{\gamma }\right)
_{\beta }\cdot \boldsymbol{\nabla }\chi _{\beta }  \label{a.22}
\end{equation}%
The transport matrix is given by (\ref{3.18}) with the correlation functions
\cite{BreyDufty01}
\begin{equation}
\mathrm{C}_{\alpha \beta }(t)=\int d\mathbf{v}\left( \mathcal{Q}^{\dag }%
\mathbf{b}_{\alpha }\right) e^{-\mathcal{L}\tau }\left( e^{\Lambda s}%
\mathcal{Q}\boldsymbol{\gamma }\right) _{\beta }.  \label{a.23}
\end{equation}%
In particular, the correlation function determining the transport
coefficient $\mu $ is
\begin{equation}
\mathrm{C}_{21}(t)=\int d\mathbf{v}\left( \mathcal{Q}^{\dag }\mathbf{b}%
_{2}\right) e^{-\mathcal{L}\tau }\left( e^{\Lambda s}\mathcal{Q}\boldsymbol{%
\gamma }\right) _{1}  \label{a.24}
\end{equation}

A more explicit form is obtained by direct solution to the equations for $%
\mathbf{F}_{1}$ and $\mathbf{F}_{2}$%
\begin{equation}
\mathbf{F}_{1}=\frac{3T}{2}\mathbf{F}_{2}-\int_{0}^{\infty }d\tau e^{-\left(
\mathcal{L}-\lambda _{1}\right) \tau }\mathcal{Q}\left( \mathbf{v}\left[
\psi _{1}-\xi \frac{3T}{4\left( \lambda _{2}-\lambda _{1}\right) }\psi _{2}%
\right] \right) .  \label{a.25}
\end{equation}%
\begin{equation}
\mathbf{F}_{2}=-\int_{0}^{\infty }d\tau e^{-\left( \mathcal{L}-\lambda
_{2}\right) \tau }\mathcal{Q}\left( \mathbf{v}\psi _{2}\left( \mathbf{v}%
\right) \right) .  \label{a.26}
\end{equation}%
The coefficient $\mu $ is then
\begin{equation}
\mu =\int_{0}^{\infty }d\tau C_{21}(\tau )\frac{\partial \chi _{1}}{\partial
n}\mid _{T}=\int d\mathbf{v}\left( \mathcal{Q}^{\dag }\mathbf{b}_{2}\right)
\left( \mathbf{F}_{1}+\mathbf{F}_{2}\frac{\partial e}{\partial n}\mid
_{T}\right) .  \label{a.27}
\end{equation}%
Substituting (\ref{a.25}) and (\ref{a.26}) leads after some rearrangement to
\begin{eqnarray}
\mu &=&-\int_{0}^{\infty }d\tau \int d\mathbf{v}\left( \mathcal{Q}^{\dag }%
\mathbf{b}_{2}\right) \left\{ e^{-\left( \mathcal{L}-\lambda _{1}\right)
\tau }\mathcal{Q}\mathbf{v}\frac{\partial f^{(0)}}{\partial n}\mid
_{T}\right.  \notag \\
&&\left. -2\frac{\partial e}{\partial n}\mid _{T}\left( e^{-\left( \mathcal{L%
}-\lambda _{2}\right) \tau }-e^{-\left( \mathcal{L}-\lambda _{1}\right) \tau
}\right) \mathcal{Q}\left( \mathbf{v}\frac{\partial f^{(0)}}{\partial e}\mid
_{n}\right) \right\}  \label{a.28}
\end{eqnarray}


\begin{thebibliography}{99}
\bibitem{ApproxMu} This is the result from reference \cite{BDKyS98}, further
simplified here by neglecting small corrections due to the distortion of the
HCS distribution from a Gaussian.
\end{thebibliography}
\end{document}